\documentclass[12pt]{article}
\usepackage{graphicx}
\usepackage{amssymb}

\begin{document}
\title{Field theories of paramagnetic Mott insulators}
\author{Subir Sachdev\\
Department of Physics, Yale University,\\ P.O. Box 208120, New
Haven CT 06520-8120}
\date{Proceedings of the International Conference on\\ Theoretical Physics, Paris, UNESCO, 22-27 July 2002.}
\maketitle

\begin{abstract}
This is a summary of a central argument in recent review articles
by the author (Physica A {\bf 313}, 252 (2002), Annals of Physics
{\bf 303}, 226 (2003), and Rev. Mod. Phys, July 2003). An
effective field theory is derived for the low energy spin singlet
excitations in a paramagnetic Mott insulator with collinear spin
correlations.
\end{abstract}

\section{INTRODUCTION}

In a recent article \cite{rmp}(intended for an audience of
experimentalists), the author has reviewed arguments that many
aspects of the physics of the cuprate superconductors can be
understood by using their proximity to paramagnetic Mott
insulators. Further, a distinction was made between Mott
insulators with collinear and non-collinear spin correlations, and
it was argued that current experimental evidence suggests that we
need only consider the collinear class. A phenomenological
description of the ground states and excitations of  these classes
of Mott insulators was provided, along with a discussion of their
experimental implications. A more technical discussion (intended
for theorists) of such insulators, along with a description of the
effective field theories which describe their low energy
properties appears in Ref.~\cite{altenberg,annals}. Here, we
briefly recall the derivation and properties of the effective
field theory of Mott insulators with {\em collinear} spin
correlations, which is expressed in terms of a compact U(1) gauge
field. The non-collinear class leads naturally to a $Z_2$ gauge
theory, but we will not consider it here. The reader is referred
to these previous reviews \cite{rmp,altenberg,annals} for complete
citations to the literature.

\section{Compact U(1) gauge theory of Mott insulators}

We focus on Mott insulators on a $d$ dimensional bipartite lattice
of sites $j$. The SU(2) spin operator ${\bf S}_j$ on site $j$ at
imaginary time $\tau$ can be written as
\begin{equation}
{\bf S}_j (\tau) = \eta_j S {\bf n} (r_j , \tau);
\end{equation}
here $\eta_j = \pm 1$ on the two sublattices, $r_j$ is the spatial
co-ordinate of site $j$, ${\bf n}$ is a unit length vector in spin
space, and $S$ is the (integer or half-odd-integer) angular
momentum of each spin. The antiferromagnetic exchange interaction
between near neighbor spins implies that ${\bf n} (r_j, \tau)$
will be a slowly varying function of its spacetime arguments. A
standard analysis of the coherent state path integral of over the
SU(2) ${\bf S}_j$ spins shows that the low energy quantum
fluctuations are described by the following partition function
\begin{eqnarray}
\mathcal{Z} &=& \int \mathcal{D} {\bf n} (r, \tau) \delta ( {\bf
n}^2 (r, \tau) - 1) \exp \Biggl[ - i S \sum_j \eta_j \int d \tau
\mathcal{A}_{\tau} ({\bf n}(r_j, \tau)) \nonumber \\
&~&~~~~~~~~~~- \frac{1}{2 g c} \int d^d r d\tau \left(
(\partial_{\tau} {\bf n})^2 + c^2 (\nabla_r {\bf n})^2 \right)
\Biggr], \label{zl}
\end{eqnarray}
where $c$ is the spin-wave velocity, and $g$ is a coupling
constant which controls the strength of the quantum fluctuations.
Excluding the first Berry phase term, this is the action of the
O(3) non-linear sigma model in $d+1$ spacetime dimensions. Here we
are primarily interested in the consequences of the Berry phases:
$\mathcal{A}_\tau ({\bf n}(\tau)) d \tau$ is defined to be the
oriented area of the spherical triangle defined by ${\bf n}
(\tau)$, ${\bf n} (\tau + d \tau)$, and an arbitrary reference
point ${\bf n}_0$ (which is usually chosen to be the north pole).

The theory (\ref{zl}) can be considered to be the ``minimal
model'' of antiferromagnets. In dimensions $d>1$ it has at least
two phases: at small $g$ there is the conventional magnetically
ordered ``N\'{e}el'' phase with $\langle {\bf n} \rangle \neq 0$,
while at large $g$ there is a ``quantum disordered'' paramagnetic
phase which preserves spin rotation invariance with $\langle {\bf
n} \rangle = 0$. We are especially interested here in the nature
of this paramagnetic state. In this section, we will manipulate
$\mathcal{Z}$ in this large $g$ regime, and derive an alternative
formulation which allows easier computation of the integral over
the Berry phases.

The key to an analysis of the large $g$ regime is a better
understanding of the nature of $\mathcal{A}_{\tau}$. We will see
that $\mathcal{A}_{\tau}$ behaves in many respects like the
time-component of a compact U(1) gauge field, and indeed, this
accounts for the suggestive notation. All physical results should
be independent of the choice of the reference point ${\bf n}_0$,
and it is easy to see by drawing triangles on the surface of a
sphere that changes in ${\bf n}_0$ amount to gauge transformations
of $\mathcal{A}_{\tau}$. If we change ${\bf n}_0$ to ${\bf
n}_0^{\prime}$, then the resulting $\mathcal{A}_{\tau}^{\prime}$
is related to $\mathcal{A}_{\tau}$ by
\begin{equation}
\mathcal{A}_{\tau}^{\prime} = \mathcal{A}_{\tau} - \partial_{\tau}
\phi (\tau) \label{gauge}
\end{equation}
where $\phi(\tau)$ measures the oriented area of the spherical
triangle defined by ${\bf n} (\tau)$, ${\bf n}_0$, and ${\bf
n}_0^{\prime}$. Furthermore, as we will discuss more completely
below, the area of any spherical triangle is uncertain modulo $4
\pi$, and this accounts for the `compactness' of the U(1) gauge
theory.

We proceed with our analysis of $\mathcal{Z}$. First, we
discretize the gradient terms of the O(3) sigma model. We will
limit our considerations here to antiferromagnets on $d$
dimensional cubic lattices, but similar considerations apply to
other bipartite lattices. We also discretize the imaginary time
direction, and (by a slight abuse of notation) use the same index
$j$ to refer to the sites of a $d+1$ dimensional cubic lattice in
spacetime. On such a lattice we can rewrite (\ref{zl}) as
\begin{equation}
Z = \int \prod_j d {\bf n}_j \delta({\bf n}_j^2 - 1) \exp \left(
\frac{1}{2g} \sum_{j,\mu} {\bf n}_j \cdot {\bf n}_{j + \hat{\mu}}
- i S \sum_j \eta_j \mathcal{A}_{j\tau} \right), \label{f1}
\end{equation}
where the sum over $\mu$ extends over the $d+1$ spacetime
directions. We have also dropped unimportant factors of the
lattice spacing and the spin-wave velocity in (\ref{f1}).

As noted above, we are especially interested here in the large $g$
regime where there are strong fluctuations of the ${\bf n}_j$.
There are strong cancellations from the Berry phases between
different spin configurations in this regime, and so the second
term in $Z$ has to be treated with great care. We will do this by
promoting the field $\mathcal{A}_{j\mu}$ to an independent degree
of freedom, while integrating out the ${\bf n}_j$. Notice that we
have now introduced all $d+1$ components of the compact U(1) gauge
field with the index $\mu$, while only the $\mu=\tau$ component
appears explicitly in (\ref{f1}). The remaining components appear
naturally as suitable degrees of freedom when we integrate the
${\bf n}_j$ out. Formally, the integration over the ${\bf n}_j$
can be done by introducing new `dummy' variables $A_{j \mu}$ and
rewriting (\ref{f1}) by introducing factors of unity on each link;
this leads to
\begin{eqnarray}
Z &=& \int \prod_{j \mu} d A_{j \mu} \exp \left(- i 2 S \sum_j
\eta_j A_{j\tau} \right) \int \prod_j d {\bf n}_j \delta({\bf
n}_j^2 - 1) \delta(\mathcal{A}_{j\mu}/2 - A_{j\mu}) \nonumber \\
&~&~~~~~~~~~~~~~~~~~~~~~~~~~~~~~~~~~~~~~~~~\times\exp \left(
\frac{1}{2g} \sum_{j,\mu} {\bf n}_j \cdot {\bf n}_{j + \hat{\mu}}
\right)
\nonumber \\
&=& \int \prod_{j \mu} d A_{j \mu} \exp \left(-\mathcal{S}_A
(A_{j\mu}) - i 2 S \sum_j \eta_j A_{j\tau} \right).
 \label{f2a}
\end{eqnarray}
In the first expression, if the integral over the $A_{j \mu}$ is
performed first, we trivially return to (\ref{f1}); however, in
the second expression we perform the integral over the ${\bf n}_j$
variables first, at the cost of introducing an unknown effective
action $\mathcal{S}_A$ for the $A_{j \mu}$. In principle,
evaluation of $\mathcal{S}_A$ may be performed order-by-order in a
``high temperature'' expansion in $1/g$: we match correlators of
the $A_{j \mu}$ flux with those of the $\mathcal{A}_{j \mu}$ flux
evaluated in the integral over the ${\bf n}_j$ with positive
weights determined only by the $1/g$ term in (\ref{f1}). Rather
than undertaking this laborious calculation, we can guess
essential features of the effective action $\mathcal{S}_A$ from
some general constraints. First, correlations in the ${\bf n}_j$
decay exponentially rapidly for large $g$ (with a correlation
length $\sim 1/\ln(g)$), and so $\mathcal{S}_A$ should be local.
Second, it should be invariant under the lattice form of the gauge
transformation (\ref{gauge})
\begin{equation}
A_{j \mu}^{\prime} = A_{j \mu} - \Delta_{\mu} \phi_j/2 \label{f3}
\end{equation}
associated with the change in the reference point on the unit
sphere from ${\bf n}_0$ to ${\bf n}_0^{\prime}$, with $\phi_j$
equal to the area of the spherical triangle formed by ${\bf n}_j$,
${\bf n}_0$ and ${\bf n}_0^{\prime}$. Finally the area of any
triangle on the sphere is uncertain modulo $4 \pi$ and so the
effective action should be invariant under
\begin{equation}
A_{j \mu} \rightarrow A_{j \mu} + 2 \pi. \label{f4}
\end{equation}
The simplest local action which is invariant under (\ref{f3}) and
(\ref{f4}) is that of {\em compact U(1) quantum electrodynamics}
and so we have
\begin{equation}
Z = \int \prod_{j\mu} d A_{j \mu} \exp \left( \frac{1}{e^2}
\sum_{\Box} \cos\left(\Delta_{\mu} A_{j \nu}- \Delta_{\nu} A_{j
\mu} \right)  - i 2S \sum_j \eta_j A_{j\tau} \right), \label{f5}
\end{equation}
for large $g$; comparison with the large $g$ expansion shows that
the coupling $e^2 \sim g^2$. In (\ref{f5}), $\Delta_{\mu}$ is the
discrete lattice derivative along the $\mu$ direction, and the sum
over $\Box$ extends over all plaquettes of the $d+1$ dimensional
cubic lattice---both notations are standard in the lattice gauge
theory literature.

The first term in the action (\ref{f5}) is, of course, the
standard `Maxwell' term of a compact U(1) gauge field. In this
language, the Berry phase has the interpretation of a $\int
J_{\mu} A_{\mu}$ coupling to a fixed matter field with `current'
$J_{\mu} = 2S\delta_{\mu\tau}$. This corresponds to static matter
with charges $\pm 2S$ on the two sublattices. It is this matter
field which will crucially control the nature of the ground state.

The remaining analysis of $Z$ depends upon the spatial
dimensionality $d$. In $d=1$, a dual model of (\ref{f5}) is
solvable, and the results are in complete accord with those
obtained earlier by Bethe ansatz and bosonization analyses of spin
chains. We will consider the $d=2$ case in the section below.
There has been relatively little discussion of the $d=3$ case
(which exhibits both confining and deconfining phases of the gauge
theory), and this remains an important avenue for future research.

\section{Duality mapping in $d=2$}
As is standard in duality mappings, we first rewrite the partition
function in $2+1$ spacetime dimensions by replacing the cosine
interaction in (\ref{f5}) by a Villain sum over periodic
Gaussians:
\begin{equation}
Z = \sum_{\{q_{\bar{\jmath}\mu}\}} \int \prod_{j\mu} d A_{j \mu}
\exp \left( -\frac{1}{2e^2} \sum_{\Box} \left(
\epsilon_{\mu\nu\lambda} \Delta_{\nu} A_{j \lambda} - 2 \pi
q_{\bar{\jmath}\mu} \right)^2 - i 2 S \sum_j \eta_j A_{j\tau}
\right), \label{f6}
\end{equation}
where $\epsilon_{\mu\nu\lambda}$ is the total antisymmetric tensor
in three dimensions, and the $q_{\bar{\jmath}\mu}$ are integers on
the links of the {\em dual} cubic lattice, which pierce the
plaquettes of the direct lattice. Throughout this subsection we
will use the index $\bar{\jmath}$ to refer to sites of this dual
lattice, while $j$ refers to the direct lattice on sites on which
the spins are located.

\begin{figure}
\centerline{\includegraphics[width=2.5in]{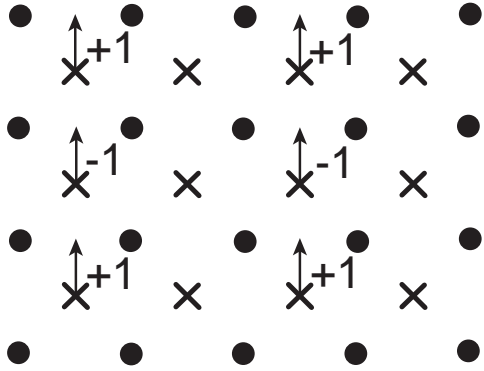}}
\caption{Specification of the non-zero values of the fixed field
$a_{\bar{\jmath}\mu}^0$. The circles are the sites of the direct
lattice, $j$, while the crosses are the sites of the dual lattice,
$\bar{\jmath}$; the latter are also offset by half a lattice
spacing in the direction out of the paper (the $\mu = \tau$
direction). The $a_{\bar{\jmath}\mu}^0$ are all zero for
$\mu=\tau,x$, while the only non-zero values of
$a_{\bar{\jmath}y}^0$ are shown above. Notice that the $a^0$ flux
obeys (\protect\ref{d2}).}\label{fig1}
\end{figure}

We will now perform a series of exact manipulations on (\ref{f6})
which will lead to a dual {\em interface} model \cite{rs,fradkiv}.
This dual model has only positive weights---this fact, of course,
makes it much more amenable to a standard statistical analysis.
This first step in the duality transformation is to rewrite
(\ref{f6}) by the Poisson summation formula:
\begin{eqnarray}
\sum_{\{q_{\bar{\jmath}\mu}\}} \exp && \left( -\frac{1}{2e^2}
\sum_{\Box} \left(\epsilon_{\mu\nu\lambda} \Delta_{\nu} A_{j
\lambda} - 2 \pi q_{\bar{\jmath}\mu} \right)^2
\right) \nonumber \\
&&~~~~~~~~~~= \sum_{\{a_{\bar{\jmath}\mu}\}} \exp \left( -
\frac{e^2}{2} \sum_{\bar{\jmath}} a_{\bar{\jmath}\mu}^2 - i
\sum_{\Box} \epsilon_{\mu\nu\lambda} a_{\bar{\jmath}\mu}
\Delta_{\nu} A_{j \lambda}\right), \label{d1}
\end{eqnarray}
where $a_{\bar{\jmath}\mu}$ (like $q_{\bar{\jmath}\mu}$) is an
integer-valued vector field on the links of the dual lattice
(here, and below, we drop overall normalization factors in front
of the partition function). Next, we write the Berry phase in a
form more amenable to duality transformations. Choose a
`background' $a_{\bar{\jmath} \mu}=a_{\bar{\jmath}}^0$ flux which
satisfies
\begin{equation}
\epsilon_{\mu\nu\lambda} \Delta_{\nu} a_{\bar{\jmath}\lambda}^0 =
\eta_j \delta_{\mu \tau}, \label{d2}
\end{equation}
where $j$ is the direct lattice site in the center of the
plaquette defined by the curl on the left-hand-side. Any
integer-valued solution of (\ref{d2}) is an acceptable choice for
$a_{\bar{\jmath}\mu}^0$, and a convenient choice is shown in
Fig~\ref{fig1}.
Using (\ref{d2}) to rewrite the Berry phase in (\ref{f6}),
applying (\ref{d1}), and shifting $a_{\bar{\jmath}\mu}$ by the
integer $2S a_{\bar{\jmath}\mu}^0$, we obtain a new exact
representation of $Z$ in (\ref{f6}):
\begin{equation}
Z = \sum_{\{ a_{\bar{\jmath} \mu} \}}  \int \prod_{j\mu} d A_{j
\mu} \exp \left( -\frac{e^2}{2} \sum_{\bar{\jmath},\mu}
(a_{\bar{\jmath}\mu}-2S a_{\bar{\jmath}\mu}^0)^2 - i \sum_{\Box}
\epsilon_{\mu\nu\lambda} a_{\bar{\jmath}\mu} \Delta_{\nu} A_{j
\lambda} \right). \label{d4}
\end{equation}
The integral over the $A_{j \mu}$ can be performed independently
on each link, and its only consequence is the imposition of the
constraint $\epsilon_{\mu\nu\lambda} \Delta_{\nu}
a_{\bar{\jmath}\lambda}=0$. We solve this constraint by writing
$a_{\bar{\jmath} \mu}$ as the gradient of a integer-valued
`height' $h_{\bar{\jmath}}$ on the sites of the dual lattice, and
so obtain
\begin{equation}
Z = \sum_{\{ h_{\bar{\jmath}} \}} \exp \left( -\frac{e^2}{2}
\sum_{\bar{\jmath},\mu} (\Delta_{\mu} h_{\bar{\jmath}}-2S
a_{\bar{\jmath}\mu}^0)^2  \right). \label{d5}
\end{equation}
This is the promised 2+1 dimensional interface, or height, model
in almost its final form.

\begin{figure}
\centerline{\includegraphics[width=4in]{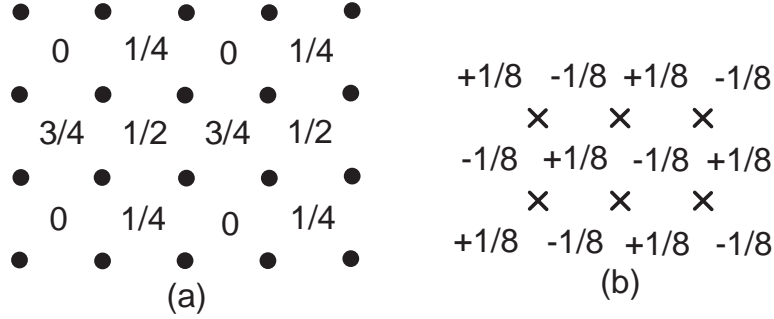}}
\caption{Specification of the non-zero values of the fixed fields
(a) $\mathcal{X}_{\bar{\jmath}}$ and (b) $\mathcal{Y}_{j \mu}$
introduced in (\protect\ref{XY}). The notational conventions are
as in Fig~\protect\ref{fig1}. Only the $\mu=\tau$ components of
$\mathcal{Y}_{j \mu}$ are non-zero, and these are shown in (b).
}\label{fig2}
\end{figure}

The physical properties of (\ref{d5}) become clearer by converting
the ``frustration'' $a_{\bar{\jmath}\mu}^0$ in (\ref{d5}) into
offsets for the allowed height values. This is done by decomposing
$a_{\bar{\jmath}\mu}^0$ into curl and divergence free parts and
writing it in terms of new fixed fields,
$\mathcal{X}_{\bar{\jmath}}$ and ${\mathcal Y}_{j \mu}$ as
follows:
\begin{equation}
a_{\bar{\jmath}\mu}^{0} = \Delta_{\mu} \mathcal{X}_{\bar{\jmath}}
+ \epsilon_{\mu\nu\lambda} \Delta_{\nu} \mathcal{Y}_{j  \lambda}.
\label{XY}
\end{equation}
The values of these new fields are shown in Fig~\ref{fig2}.
Inserting (\ref{XY}) into (\ref{d5}), we can now write the height
model in its simplest form \cite{rs}
\begin{equation}
Z_h = \sum_{\{H_{\bar{\jmath}}\}} \exp \left ( - \frac{e^2}{2}
\sum_{\bar{\jmath}} \left( \Delta_{\mu} H_{\bar{\jmath}} \right)^2
\right), \label{he1}
\end{equation}
where
\begin{equation}
H_{\bar{\jmath}} \equiv h_{\bar{\jmath}} - 2 S
\mathcal{X}_{\bar{\jmath}} \label{he2}
\end{equation}
is the new height variable we shall work with. Notice that the
$\mathcal{Y}_{j \mu}$ have dropped out, while the
$\mathcal{X}_{\bar{\jmath}}$ act only as fractional offsets (for
$S$ not an even integer) to the integer heights. From (\ref{he2})
we see that for half-odd-integer $S$ the height is restricted to
be an integer on one of the four sublattices, an integer plus 1/4
on the second, an integer plus 1/2 on the third, and an integer
plus 3/4 on the fourth; the fractional parts of these heights are
as shown in Fig~\ref{fig2}a; the steps between neighboring heights
are always an integer plus 1/4, or an integer plus 3/4. For $S$ an
odd integer, the heights are integers on one square sublattice,
and half-odd-integers on the second sublattice. Finally for even
integer $S$ the offset has no effect and the height is an integer
on all sites. We discuss these classes of $S$ values in turn in
the following subsections.

\subsection{$S$ even integer}
\label{S2}

In this case the offsets $2S \mathcal{X}_{\bar{\jmath}}$ are all
integers, and (\ref{he1}) is just an ordinary three dimensional
height model which has been much studied in the literature. Unlike
the two-dimensional case, three-dimensional height models
generically have no roughening transition, and the interface is
always smooth. With all heights integers, the smooth phase breaks
no lattice symmetries. So square lattice antiferromagnets with $S$
even integer can have a paramagnetic ground state with a spin gap
and no broken symmetries. This is in accord with the exact ground
state for a $S=2$ antiferromagnet on the square lattice found by
Affleck {\em et al.}, the AKLT state \cite{aklt}.

\subsection{$S$ half-odd-integer}
\label{Shint}

Now the heights of the interface model can take four possible
values, which are integers plus the offsets on the four square
sublattices shown in Fig~\ref{fig2}a. As in Section~\ref{S2}, the
interface is always smooth {\em i.e.} any state of (\ref{he1}) has
a fixed average interface height $\sum_{\bar{\jmath}} \langle
H_{\bar{\jmath}} \rangle$, and {\em any} well-defined value for
this average height breaks the uniform shift symmetry of the
height model under which $H_{\bar{\jmath}} \rightarrow
H_{\bar{\jmath}} \pm 1$. After accounting for the height offsets,
we will see below that any smooth interface must also break a
lattice symmetry with the development of {\em bond order}: this
allows a number of distinct spin gap ground states of the lattice
antiferromagnet.

\begin{figure}
\centerline{\includegraphics[width=2.3in]{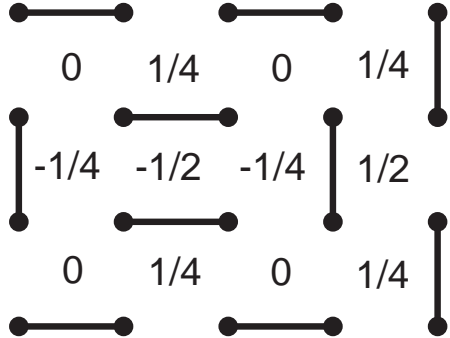}}
\caption{Mapping between the quantum dimer model and the interface
model $Z$ in (\protect\ref{he1}). Each dimer on the direct lattice
is associated with a step in height of $\pm 3/4$ on the link of
the dual lattice that crosses it. All other height steps are $\pm
1/4$. Each dimer represents a singlet valence between the sites,
as in Fig~\protect\ref{fig2}.}\label{fig3}
\end{figure}

It is useful, first, to obtain a simple physical interpretation of
the interface model in the language of the $S=1/2$ antiferromagnet
\cite{zheng}. From Fig~\ref{fig2}a it is clear that nearest
neighbor heights can differ either by 1/4 or 3/4 (modulo
integers). To minimize the action in (\ref{he1}), we should choose
the interface with the largest possible number of steps of $\pm
1/4$. However, the interface is frustrated, and it is not possible
to make all steps $\pm 1/4$ and at least a quarter of the steps
must be $\pm 3/4$. Indeed, there is a precise one-to-one mapping
between interfaces with the minimal number of $\pm 3/4$ steps (we
regard interfaces differing by a uniform integer shift in all
heights as equivalent) and the dimer coverings of the square
lattice: the proof of this claim is illustrated in Fig~\ref{fig3}.
We identify each dimer with a singlet valence bond between the
spins (the ellipses in Fig~\ref{fig2}), and so each interface
corresponds to a quantum state with each spin locked in the a
singlet valence bond with a particular nearest neighbor.
Fluctuations of the interface in imaginary time between such
configurations correspond to quantum tunneling events between such
dimer states, and an effective Hamiltonian for this is provided by
the quantum dimer model \cite{qd}.

\begin{figure}
\centerline{\includegraphics[width=2.7in]{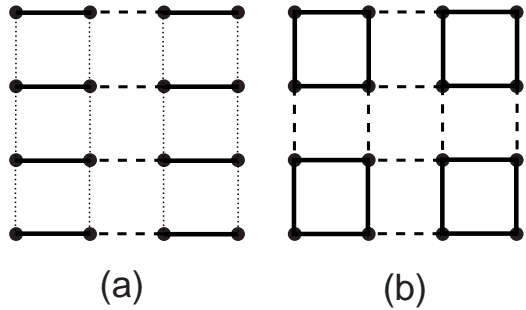}}
\caption{Sketch of the two simplest possible states with bond
order for $S=1/2$ on the square lattice: (a) the columnar
spin-Peierls states, and (b) plaquette state. The different values
of the $\langle {\bf S}_i \cdot {\bf S}_j \rangle$ on the links
are encoded by the different line styles. Both states are 4-fold
degenerate; an 8-fold degenerate state, with superposition of the
above orders, also appears as a possible ground state of the
generalized interface model.} \label{fig4}
\end{figure}

The nature of the possible smooth phases of the interface model
are easy to determine from the above picture and by standard
techniques from statistical theory \cite{rs,zheng}. Interfaces
with average height $\langle H_{\bar{\jmath}} \rangle =
1/8,3/8,5/8,7/8$ (modulo integers) correspond to the four-fold
degenerate bond-ordered states in Fig~\ref{fig4}a, while those
with $\langle H_{\bar{\jmath}} \rangle = 0,1/4,1/2,3/4$ (modulo
integers) correspond to the four-fold degenerate plaquette
bond-ordered states in Fig~\ref{fig4}b.
All other values of $\langle H_{\bar{\jmath}} \rangle$ are
associated with eight-fold degenerate bond-ordered states with a
superposition of the orders in Fig~\ref{fig4}a and b.

Support for the class of bond-ordered states described above has
appeared in a number of numerical studies of $S=1/2$
antiferromagnets in $d=2$ which have succeeded in moving from the
small $g$ N\'{e}el phase to the large $g$ paramagnet. These
include studies on the honeycomb lattice \cite{fouet1}, on the
planar pyrochlore lattice \cite{fouet2}, on square lattice models
with ring-exchange and easy-plane spin symmetry \cite{sandvik},
and square lattice models with SU($N$) symmetry \cite{harada}.

\subsection{$S$ odd integer}

This case is similar to that $S$ half-odd-integer, and we will not
consider it in detail. The Berry phases again induce bond order in
the spin gap state, but this order need only lead to a two-fold
degeneracy.

\section{Conclusions}

The primary topic discussed in this paper has been the effective
field theory of paramagnetic Mott insulators with collinear spin
correlations. This field theory is the compact U(1) gauge theory
in (\ref{f5}), and applies in all spatial dimensions. We also
reviewed duality mappings of (\ref{f5}) which are special to $d=2$
spatial dimensions, and mapped the theory onto the interface model
(\ref{he1}). Finally, we reiterate that paramagnetic Mott
insulators with non-collinear spin correlations are described by a
$Z_2$ gauge theory which has not been presented here.

\end{document}